\documentclass[a4paper,12pt]{article}
\usepackage{amssymb}
\usepackage{latexsym}
\usepackage[dvips]{graphicx}
\usepackage{cite}
\newcount\hour
\newcount\minute
\newtoks\amorpm
\hour=\time\divide\hour by60 \minute=\time{\multiply\hour by60 \global\advance\minute by-\hour}
\edef\standardtime{{\ifnum\hour<12 \global\amorpm={am}%
        \else\global\amorpm={pm}\advance\hour by-12 \fi
        \ifnum\hour=0 \hour=12 \fi
        \number\hour:\ifnum\minute<10 0\fi\number\minute\the\amorpm}}
\edef\militarytime{\number\hour:\ifnum\minute<10 0\fi\number\minute}
\newcommand{\be}{\begin{equation}}
\newcommand{\ee}{\end{equation}}
\newcommand{\bea}{\begin{eqnarray}}
\newcommand{\nn}{\nonumber}
\newcommand{\eea}{\end{eqnarray}}

\begin{document}

\begin{titlepage}
\begin{flushright}
UB-ECM-PF-03/20
\end{flushright}
\begin{centering}
\vspace{.3in}
{\Large{\bf Energy  and Momentum Distributions\\ of a ${\bf(2+1)}$-dimensional black hole background}}
\\

\vspace{.5in} {\bf  Elias C.
Vagenas\footnote{evagenas@ecm.ub.es} }\\

\vspace{0.3in}

Departament d'Estructura i Constituents de la Mat\`{e}ria\\
and\\ CER for Astrophysics, Particle Physics and Cosmology\\
Universitat de Barcelona\\
Av. Diagonal 647\\ E-08028 Barcelona\\
Spain\\
\end{centering}

\vspace{0.7in}
\begin{abstract}
Using Einstein, Landau-Lifshitz, Papapetrou and Weinberg  energy-momentum complexes we explicitly evaluate the
energy and momentum distributions associated with a non-static and circularly symmetric three-dimensional
spacetime. The gravitational background under study is an exact solution of the Einstein's equations in the
presence of a cosmological constant and a null fluid. It can be regarded as the three-dimensional analogue of the
Vaidya metric and represents a non-static spinless ($2+1$)-dimensional black hole with an outflux of null
radiation. All four above-mentioned prescriptions give exactly the same energy and momentum distributions for the
specific black hole background. Therefore, the results obtained here provide evidence in support of the claim that
for a given gravitational background, different energy-momentum complexes can give identical results in three
dimensions. Furthermore, in the limit of zero cosmological constant the results presented here reproduce the
results obtained by Virbhadra who utilized the Landau-Lifshitz energy-momentum complex for the same
($2+1$)-dimensional black hole background in the absence of a
cosmological constant.\\
 \end{abstract}

\end{titlepage}
\newpage

\baselineskip=18pt
\section*{Introduction}
Energy-momentum localization has been one of the oldest, most interesting but also most controversial problems
in gravitation. Many renowned physicists have been working on this problematic issue with
Einstein to be first in the row. After Einstein's seminal work \cite{einstein} on energy-momentum complexes a
large number of expressions for the energy distribution were proposed \cite{ll}-\hspace{-0.1ex}\cite{weinberg}.
 However, the idea of the energy-momentum complex was severely criticized for a number of reasons
\cite{chandra}-\hspace{-0.1ex}\cite{chiang}. Therefore, this approach was abandoned for a long time.
In 1990 Virbhadra  revived the interest in this approach
\cite{vir1,vir2,vir3} and since then numerous works on evaluating the energy and momentum distributions
of several gravitational
backgrounds have been completed \cite{par1}-\hspace{-0.1ex}\cite{vir4}.
Later attempts to deal with this problematic issue were made
 by proposers of quasi-local approach. The seminal work on quasilocal mass was that of
Brown and York  \cite{brown1}. However, it should be stressed that the determination as well as the computation of
the quasilocal energy and quasilocal angular momentum of a ($2+1$)-dimensional gravitational background were first
presented by Brown, Creighton and Mann \cite{mann1}. Many attempts since then have been performed to give new
definitions of quasilocal energy in General Relativity \cite{sean}-\hspace{-0.1ex}\cite{yau2}. Considerable
efforts  have also been performed in constructing superenergy tensors \cite{senovilla1}. Motivated by the works of
Bel \cite{bel1}-\hspace{-0.1ex}\cite{bel3} and independently of Robinson \cite{robinson}, many investigations have
been given in this field \cite{senovilla2}--\hspace{-0.1ex}\cite{senovilla4}.
\par\noindent
In this paper we are implementing the approach of energy momentum complexes
\footnote{Recently, there was some interest in  using the approach of the energy-momentum complexes in the
framework of teleparallel equivalent of General Relativity, i.e. teleparallel gravity \cite{andra1,vargas}
(for a recent review on teleparallel gravity see \cite{andra2})}.
The gravitational background  under investigation is a three-dimensional black hole spacetime.
Our interest in three-dimensional gravitational backgrounds
stems to the fact that ($2+1$) dimensions provide a simpler framework than ($3+1$) and
a more realistic one than ($1+1$) to investigate various problems.
We evaluate the energy and momentum distributions associated with the
($2+1$)-dimensional black hole using four different
energy-momentum complexes, specifically we are implementing the prescriptions of
Einstein, Landau-Lifshitz, Papapetrou and Weinberg.
The specific ($2+1$)-dimensional black hole background is a non-static although spinless (c.f. spinless BTZ
black hole solution\cite{jorge}), and
circularly symmetric \cite{shwetket}. It  is an exact solution of the Einstein equations in the presence
 of a cosmological constant $\Lambda$ and a null fluid. Therefore, it can be regarded as the three-dimensional analogue
 of the Vaidya metric and represents a non-static spinless ($2+1$)-dimensional black hole with an outflux of null
 radiation.
\par\noindent
The remainder of the paper is organized as follows.
In Section 1 we briefly present the $(2+1)$-dimensional black hole spacetime
in which energy and momentum distributions by using four different prescriptions
are to be calculated.
In the subsequent four Sections using Einstein, Landau-Lifshitz, Papapetrou and Weinberg
energy-momentum complexes, respectively, we explicitly evaluate the energy and momentum distributions contained
in a ``sphere"of fixed radius.
The results extracted in the four different prescriptions
associated with the same gravitational background are identical.
 Finally, Section 6 is devoted to a brief summary of results and concluding remarks.
\section{The non-static ($\bf{2+1}$)-dimensional black hole}
In 1992 Banados, Teitelboim and Zanelli discovered a black hole solution (known as  BTZ black hole) in ($2+1$)
dimensions  \cite{jorge}. Till that time it was believed that no black hole solution exists in three-dimensional
spacetimes \cite{abbott}. Since this discovery, many investigations have been performed in order to extend the BTZ
black hole solution. Virbhadra found one of these extended solutions \cite{shwetket} by including a cosmological
constant $\Lambda$ and an energy-momentum tensor of a null fluid, in the Einstein's field equations\footnote{A
classification of solutions of the three-dimensional Einstein equations was given by Yamazaki and Ida in
\cite{yamazaki}.}, i.e. \be R_{\mu\nu}-\frac{1}{2}g_{\mu\nu}R+g_{\mu\nu}\Lambda = \kappa\,T_{\mu\nu}^{fluid}
\label{fieldeqns} \ee where $\kappa$ is the gravitational coupling constant.
\par\noindent
The energy-momentum of the null fluid $T_{\mu\nu}^{fluid}$ is
given by \be T_{\mu\nu}^{fluid}=U_{\mu}U_{\nu} \ee where $U_{\mu}$ is the fluid current vector satisfying the equation
\be U_{\mu}U^{\mu}=0 \ee since the fluid is null.
\par\noindent
An exact solution of the field equations (\ref{fieldeqns}) is given by the line element \be
ds^{2}=-\left(-m(u)-\Lambda r^{2}\right)du^{2} - 2du\,dr + r^{2}d\phi^{2} \label{metric1} \ee where $u$ is the
retarded null coordinate ($u=t-r$). This solution\footnote{It is noteworthy that at the same period of time Chan,
Chan and Mann \cite{mann}, and Husain \cite{husain} trying to investigate the phenomenon of mass inflation
discovered non-static ($2+1$) solutions describing black holes irradiated by  an influx of null radiation.} is
obviously non-static and circularly symmetric, and thus there is only one killing vector $\partial/\partial\phi$
related to the rotational invariance.
\par\noindent
Therefore, solution (\ref{metric1}) can be regarded as the three-dimensional analogue
 of the Vaidya metric and represents a non-static ($2+1$)-dimensional black hole with an outflux of null
 radiation. In the case where the mass function $m(u)$ is constant the black hole under consideration becomes the
 spinless BTZ black hole.
\section{Einstein's Prescription}
The energy-momentum complex of Einstein \cite{einstein} in a  three-dimensional background is given as
\footnote{It should be noted that it was actually von Freud who showed that the Einstein energy-stress complex can
be written as the divergence of an antisymmetric superpotential \cite{freud}.}
\be
\theta^{\mu}_{\nu}=\frac{1}{2\kappa}h^{\mu\lambda}_{\nu\hspace{1ex},\lambda}
\label{etheta}
\ee
where the Einstein's superpotential $ h^{\mu\lambda}_{\nu}$ is of the form
\be
h^{\mu\lambda}_{\nu}=\frac{1}{\sqrt{-g}} g_{\nu\sigma}\left[-g\,\left(g^{\mu\sigma}g^{\lambda\kappa}\,
-\,g^{\lambda\sigma}g^{\mu\kappa}\right)\right]_{,\kappa}
\label{esuper}
\ee
with the antisymmetric property
\be
h^{\mu\lambda}_{\nu}=-h^{\lambda\mu}_{\nu} \hspace{1ex}.
\ee Thus, the energy and momentum in Einstein's
prescription for a three-dimensional background are given by
\be
P_{\mu}=\int\int\int
\theta^{t}_{\mu}dx^{1}dx^{2}
\label{emomentum}
\ee
and specifically the energy of the physical system in a three-dimensional
background is
\be
E=\int\int\int \theta^{t}_{t}dx^{1}dx^{2}\hspace{1ex}.
\label{eenergy}
\ee
It should be noted that the calculations have to be restricted
to the use of quasi-Cartesian coordinates.
\par\noindent
In order to evaluate the energy and momentum distributions in the non-static ($2+1$)-dimensional black hole
background described by the line element (\ref{metric1}), we firstly have to calculate the Einstein's
superpotentials. There are sixteen non-vanishing superpotentials in Einstein's prescription
for the gravitational background under study but the required ones are the following
\bea
h^{tt}_{t}&=&h^{tt}_{x}=h^{tt}_{y}=0\,,\nn\\
h^{tx}_{t}&=&-\frac{x}{r^{2}}\left(1+\Lambda r^{2}+m\right)\,,\nn\\
h^{ty}_{t}&=&-\frac{y}{r^{2}}\left(1+\Lambda r^{2}+m\right)\,,\nn\\
h^{tx}_{x}&=&\frac{x^{2}+\Lambda\left(x^4 + 3x^{2}y^{2}+2y^{4}\right) +x^{2}m+y^{2}r\left(\dot{m}+m'\right)}{r^{3}}\,\label{einsuper}\\
h^{ty}_{x}&=&-\frac{xy\left(-1+\Lambda\left(x^2 + y^{2}\right) - m + r\left(\dot{m}+m'\right)\right)}{r^{3}}\,,\nn\\
h^{tx}_{y}&=&-\frac{xy\left(-1+\Lambda\left(x^2 + y^{2}\right) - m + r\left(\dot{m}+m'\right)\right)}{r^{3}}\,,\nn\\
h^{ty}_{y}&=&\frac{y^{2}+\Lambda\left(y^4 + 3x^{2}y^{2}+2x^{4}\right) +y^{2}m+x^{2}r\left(\dot{m}+m'\right)}{r^{3}}\,,\nn
\eea
where the dot and prime denote the partial derivatives with respect to time $t$ and radial coordinate $r$,
respectively. The radial coordinate in terms of the quasi-Cartesian
coordinates  is given as
\be
r=\sqrt{x^{2}+y^{2}}
\ee
while  the mass parameter $m=m(u)$ is given by
\be
m(u)=m(t,\sqrt{x^{2}+y^{2}})\hspace{1ex}.
\ee
By substituting the Einstein's superpotentials as given by equation (\ref{einsuper}) into equation (\ref{etheta})
one gets the energy density distribution
\be
\theta^{t}_{t}=\frac{m'}{2\kappa r}+\frac{\Lambda}{\kappa}
\label{eindensity}
\ee
while the momentum density distributions are given by
\bea
\theta^{t}_{x}&=&-\frac{x\dot{m}}{2\kappa r^{2}}\nn\\
\theta^{t}_{x}&=&-\frac{y\dot{m}}{2\kappa r^{2}}\hspace{1ex}.
\label{einmomendensity}
\eea
Therefore, if we substitute expressions (\ref{eindensity}) and (\ref{einmomendensity}) into equations
(\ref{eenergy}) and (\ref{emomentum}), respectively, the energy and momentum distributions  associated
with the non-static ($2+1$)-dimensional black hole under study, which are
contained in a ``sphere''\footnote{Since the spatial section of the ($2+1$) spacetime is two-dimensional
the ``sphere'' is just a circle.}  of radius $r_{0}$,  are given by
\bea
E(r_{0})&=&\frac{\pi}{\kappa}\left(m + \Lambda r^{2}_{0}\right)\nn\\
P_{x}&=&0\label{eindistrib}\\
P_{y}&=&0\hspace{1ex}.\nn
\eea
A couple of comments are in order. Firstly, a neutral test particle experiences at a finite distance $r_{0}$
the gravitational field of the effective gravitational mass given by the first of the expressions
in (\ref{eindistrib}).  What is worth mentioning is the fact that if the cosmological constant is negative, i.e.
\be
\Lambda=-\frac{1}{l^2}\hspace{1ex},
\ee
then  it seems that it is possible, although in a gravitational field, the neutral particle to
move freely on the circle of radius $r_0$ whenever the condition
\be
m(t,r_{0})=\frac{r^{2}_{0}}{l^{2}}
\label{condition}
\ee
is fulfilled. Secondly, it easily seen that the energy-momentum complex of Einstein as
formulated in the gravitational background under consideration satisfies the local conservation laws, i.e
the conservation laws in the ordinary sense,
\be
\frac{\partial}{\partial x^{\mu}}\,\theta^{\mu}_{\nu}=0\hspace{1ex}.
\ee
\section{Landau-Lifshitz's Prescription}
The energy-momentum complex of Landau-Lifshitz \cite{ll} in a  three-dimensional background is given by
\be
L^{\mu\nu}=\frac{1}{2\kappa}S^{\mu\kappa\nu\lambda}_{\hspace{4ex},\kappa\lambda}
\label{lltheta}
\ee
where the Landau-Lifshitz's superpotential $S^{\mu\kappa\nu\lambda}$ is of the form
\be
S^{\mu\kappa\nu\lambda}=-g\left(g^{\mu\nu}g^{\kappa\lambda}-g^{\mu\lambda}g^{\kappa\nu}\right)
\hspace{1ex}.
\label{llsuper}
\ee
The  energy-momentum complex of Landau-Lifshitz is symmetric in its indices
\be
L^{\mu\nu}=L^{\nu\mu} \hspace{1ex}.
\ee
The energy and momentum in Landau-Lifshitz's
prescription for a three-dimensional background are given by
\be
P^{\mu}=\int\int\int
L^{t \mu}dx^{1}dx^{2}
\label{llmomentum}
\ee
and specifically the energy of the physical system in a three-dimensional
background is
\be
E=\int\int\int L^{tt}dx^{1}dx^{2}\hspace{1ex}.
\label{llenergy}
\ee
It should be noted that the calculations in the Landau-Lifshitz's prescription have to be restricted
to the use of quasi-Cartesian coordinates as in the Einstein's prescription.
\par\noindent
Since we intend to evaluate the energy and momentum distributions in the non-static ($2+1$)-dimensional black hole
background described by the line element (\ref{metric1}), we firstly calculate the Landau-Lifshitz's
superpotentials. There are thirty six  non-vanishing superpotentials in Landau-Lifshitz's prescription
for the specific gravitational background given by (\ref{metric1}) but the required ones are the following
\bea
S^{tttt}&=&S^{tttx}=S^{ttty}=S^{txtt}=S^{tytt}=S^{xttt}=0\,,\nn\\
S^{txtx}&=&-\frac{x^{2}+2y^{2}+\Lambda\left(x^{2}y^{2}+y^{4}\right)+y^{2}m}{r^2}\,,\nn\\
S^{txty}&=&\frac{xy\left(1+\Lambda\left(x^{2}+y^{2}\right)+m\right)}{r^2}\,,\nn\\
S^{tytx}&=&\frac{xy\left(1+\Lambda\left(x^{2}+y^{2}\right)+m\right)}{r^2}\,,\nn\\
S^{tyty}&=&-\frac{y^{2}+2x^{2}+\Lambda\left(x^{2}y^{2}+x^{4}\right)+x^{2}m}{r^2}\,,\nn\\
S^{xttx}&=&\frac{x^{2}+2y^{2}+\Lambda\left(x^{2}y^{2}+y^{4}\right)+y^{2}m}{r^2}\,,\nn\\
S^{xtty}&=&-\frac{xy\left(1+\Lambda\left(x^{2}+y^{2}\right)+m\right)}{r^2}\,,\nn\\
S^{xxtt}&=&S^{xxtx}=S^{xxty}=S^{xytt}=S^{yttt}=S^{yxtt}=0\,,\label{llsuper}\\
S^{xytx}&=&\frac{y\left(1+\Lambda\left(x^{2}+y^{2}\right)+m\right)}{r}\,,\nn\\
S^{xyty}&=&-\frac{x\left(1+\Lambda\left(x^{2}+y^{2}\right)+m\right)}{r}\,,\nn\\
S^{yttx}&=&-\frac{xy\left(1+\Lambda\left(x^{2}+y^{2}\right)+m\right)}{r^2}\,,\nn\\
S^{ytty}&=&\frac{y^{2}+2x^{2}+\Lambda\left(x^{2}y^{2}+x^{4}\right)+x^{2}m}{r^2}\,,\nn\\
S^{yxtx}&=&-\frac{y\left(1+\Lambda\left(x^{2}+y^{2}\right)+m\right)}{r}\,,\nn\\
S^{yxty}&=&\frac{x\left(1+\Lambda\left(x^{2}+y^{2}\right)+m\right)}{r}\,,\nn\\
S^{yytt}&=&S^{yytx}=S^{yyty}=0\hspace{1ex}.\nn\\
\eea
 The mass parameter $m=m(u)$ in terms of the quasi-Cartesian coordinates is given again by
\be
m(u)=m(t,\sqrt{x^{2}+y^{2}})\hspace{1ex}.
\ee
By substituting the Landau-Lifshitz's superpotentials as given in equation (\ref{llsuper}) into equation (\ref{lltheta})
one gets the energy density distribution
\be
L^{tt}=\frac{m'}{2\kappa r}+\frac{\Lambda}{\kappa}
\label{lldensity}
\ee
while the momentum density distributions are given by
\bea
L^{tx}&=&-\frac{x\dot{m}}{2\kappa r^{2}}\nn\\
L^{ty}&=&-\frac{y\dot{m}}{2\kappa r^{2}}\hspace{1ex}.
\label{llmomendensity}
\eea
Therefore, if we substitute expressions (\ref{lldensity}) and (\ref{llmomendensity}) into equations
(\ref{llenergy}) and (\ref{llmomentum}), respectively, the energy and momentum distributions
associated with the non-static ($2+1$)-dimensional black hole under study, which are contained in
a ``sphere''  of radius $r_{0}$,  are given by
\bea
E(r_{0})&=&\frac{\pi}{\kappa}\left(m + \Lambda r^{2}_{0}\right)\nn\\
P_{x}&=&0\label{lldistrib}\\
P_{y}&=&0\hspace{1ex}.\nn
\eea
It is clear that we have derived exactly the same energy and momentum distributions as in the case of the
Einstein's prescription (c.f. equations (\ref{eindensity}), (\ref{einmomendensity}) and (\ref{eindistrib})).
The comment concerning the neutral test particle and the effective gravitational
mass made in the previous section also holds here.
\par\noindent
Furthermore, the  energy-momentum complex of Landau-Lifshitz
as formulated in the ($2+1$)-dimensional black hole under investigation, satisfies the local
conservation laws
\be
\frac{\partial}{\partial x^{\nu}}\,L^{\mu\nu}=0
\hspace{1ex}.
\ee
\section{Papapetrou's Prescription}
The energy-momentum complex of Papapetrou \cite{pp} in a  three-dimensional background is given by
\be
\Sigma^{\mu\nu}=\frac{1}{2\kappa}N^{\mu\nu\kappa\lambda}_{\hspace{4ex},\kappa\lambda}
\label{pptheta}
\ee
where the Papapetrou's superpotential $N^{\mu\nu\kappa\lambda}$ is given by
\be
N^{\mu\nu\kappa\lambda}=\sqrt{-g}\left(g^{\mu\nu}g\eta^{\kappa\lambda}-g^{\mu\kappa}\eta^{\nu\lambda}
+g^{\kappa\lambda}\eta^{\mu\nu} -g^{\nu\lambda}\eta^{\mu\kappa}\right)
\label{ppsuper}
\ee
where $\eta^{\mu\nu}$ is the Minkowskian metric, i.e.
\be
(\eta^{\mu\nu})=\left[
\begin{array}{rrr}
1&0&0\\
0&-1&0\\
0&0&-1\\
\end{array}\right]
\label{minkowski}
\hspace{1ex}.
\ee
The  energy-momentum complex of Papapetrou like the Landau-Lifshitz one, is symmetric in its indices
\be
\Sigma^{\mu\nu}=\Sigma^{\nu\mu} \hspace{1ex}.
\ee
The energy and momentum in Papapetrou's
prescription for a three-dimensional background are given by
\be
P^{\nu}=\int\int\int
\Sigma^{t \nu}dx^{1}dx^{2}
\label{ppmomentum}
\ee
and specifically the energy of the physical system in a three-dimensional
background is
\be
E=\int\int\int \Sigma^{tt}dx^{1}dx^{2}\hspace{1ex}.
\label{ppenergy}
\ee
It should be noted that the calculations in the Papapetrou's prescription have to be restricted
to the use of quasi-Cartesian coordinates as in the two afore-said prescriptions.
\par\noindent
Since our scope is to evaluate the energy and momentum distributions in the non-static ($2+1$)-dimensional black hole
background described by the line element (\ref{metric1}) we firstly calculate the Papapetrou's
superpotentials. There are fifty four  non-vanishing superpotentials in Papapetrou's prescription
for the specific gravitational background given by (\ref{metric1}) but the required ones are the following
\bea
N^{tttt}&=&0\,,\nn\\
N^{tttx}&=&-\frac{x\left(1+\Lambda\left(x^{2}+y^{2}\right)+m\right)}{r}\,,\nn\\
N^{ttty}&=&-\frac{y\left(1+\Lambda\left(x^{2}+y^{2}\right)+m\right)}{r}\,,\nn\\
N^{ttxt}&=&\frac{x\left(1+\Lambda\left(x^{2}+y^{2}\right)+m\right)}{r}\,,\nn\\
N^{ttxx}&=&\frac{2x^{2}+3y^{2}+\Lambda\left(x^{2}y^{2}+y^{4}\right)+y^{2}m}{r^2}\,,\nn\\
N^{ttxy}&=&-\frac{xy\left(1+\Lambda\left(x^{2}+y^{2}\right)+m\right)}{r^2}\,,\nn\\
N^{ttyt}&=&\frac{y\left(1+\Lambda\left(x^{2}+y^{2}\right)+m\right)}{r}\,,\nn\\
N^{ttyx}&=&-\frac{xy\left(1+\Lambda\left(x^{2}+y^{2}\right)+m\right)}{r^2}\,,\nn\\
N^{ttyy}&=&\frac{2y^{2}+3x^{2}+\Lambda\left(x^{2}y^{2}+x^{4}\right)+x^{2}m}{r^2}\,,\nn\\
N^{xttt}&=&0\,,\nn\\
N^{xttx}&=&-\left(2+\Lambda\left(x^{2}+y^{2}\right)+m\right)\,,\nn\\
N^{xtty}&=&0\,,\label{ppsuper}\\
N^{xtxt}&=&\frac{(-y^{2})+\Lambda\left(x^{2}y^{2}+x^{4}\right)+x^{2}m}{r^2}\,,\nn\\
N^{xtxx}&=&N^{xtxy}=0\,,\nn\\
N^{xtyt}&=&\frac{xy\left(1+\Lambda\left(x^{2}+y^{2}\right)+m\right)}{r^2}\,,\nn\\
N^{xtyx}&=&-\frac{y\left(1+\Lambda\left(x^{2}+y^{2}\right)+m\right)}{r}\,,\nn\\
N^{xtyy}&=&\frac{x\left(1+\Lambda\left(x^{2}+y^{2}\right)+m\right)}{r}\,,\nn\\
N^{yttt}&=&N^{yttx}=0\,,\nn\\
N^{ytty}&=&-\left(2+\Lambda\left(x^{2}+y^{2}\right)+m\right)\,,\nn\\
N^{ytxt}&=&\frac{xy\left(1+\Lambda\left(x^{2}+y^{2}\right)+m\right)}{r^2}\,,\nn\\
N^{ytxx}&=&\frac{y\left(1+\Lambda\left(x^{2}+y^{2}\right)+m\right)}{r}\,,\nn\\
N^{ytxy}&=&-\frac{x\left(1+\Lambda\left(x^{2}+y^{2}\right)+m\right)}{r}\,,\nn\\
N^{ytyt}&=&\frac{(-x^{2})+\Lambda\left(x^{2}y^{2}+y^{4}\right)+y^{2}m}{r^2}\,,\nn\\
N^{ytyx}&=&N^{ytyy}=0\hspace{1ex}.\nn
\eea
The mass parameter $m=m(u)$ in terms of the quasi-Cartesian
coordinates is given as before by
\be
m(u)=m(t,\sqrt{x^{2}+y^{2}})\hspace{1ex}.
\ee
By substituting the Papapetrou's superpotentials (\ref{ppsuper}) into equation (\ref{pptheta})
one gets the energy density distribution
\be
\Sigma^{tt}=\frac{m'}{2\kappa r}+\frac{\Lambda}{\kappa}
\label{ppdensity}
\ee
while the momentum density distributions are given by
\bea
\Sigma^{tx}&=&-\frac{x\dot{m}}{2\kappa r^{2}}\nn\\
\Sigma^{ty}&=&-\frac{y\dot{m}}{2\kappa r^{2}}\hspace{1ex}.
\label{ppmomendensity}
\eea
Therefore, if we substitute expressions (\ref{ppdensity}) and (\ref{ppmomendensity}) into equations
(\ref{ppenergy}) and (\ref{ppmomentum}), respectively, respectively, the energy and momentum distributions
associated with the non-static ($2+1$)-dimensional black hole under investigation, which are contained in
a ``sphere''  of radius $r_{0}$,  are given by
\bea
E(r_{0})&=&\frac{\pi}{\kappa}\left(m + \Lambda r^{2}_{0}\right)\nn\\
P_{x}&=&0\label{ppdistrib}\\
P_{y}&=&0\hspace{1ex}.\nn
\eea
It is obvious that we have again derived exactly the same energy and momentum distributions as in the cases of the
Einstein's and Landau-Lifshitz's prescriptions.
The comment concerning the neutral test particle and the effective gravitational
mass made in Section 1 also holds here.
\par\noindent
Furthermore, the  energy-momentum complex of Papapetrou
as formulated in the ($2+1$)-dimensional black hole under study, satisfies the local
conservation laws
\be
\frac{\partial}{\partial x^{\nu}}\,\Sigma^{\mu\nu}=0
\hspace{1ex}.
\ee
\section{Weinberg's Prescription}
The energy-momentum complex of Weinberg in a  three-dimensional background is given by \cite{weinberg}
\be
\tau^{\nu\lambda}=\frac{1}{2\kappa}Q^{\rho\nu\lambda}_{\hspace{3ex},\rho}
\label{weintheta}
\ee
where the Weinberg's superpotential $Q^{\rho\nu\lambda}$ is given by
\be
Q^{\rho\nu\lambda}=\frac{\partial h^{\mu}_{\mu}}{\partial x_{\nu}}\eta^{\rho\lambda}-
\frac{\partial h^{\mu}_{\mu}}{\partial x_{\rho}}\eta^{\nu\lambda}-
\frac{\partial h^{\mu\nu}}{\partial x^{\mu}}\eta^{\rho\lambda}+
\frac{\partial h^{\mu\rho}}{\partial x^{\mu}}\eta^{\nu\lambda}+
\frac{\partial h^{\nu\lambda}}{\partial x_{\rho}}-
\frac{\partial h^{\rho\lambda}}{\partial x_{\nu}}
\label{weinsuper}
\ee
where the symmetric tensor $h_{\mu\nu}$ is
\be
h_{\mu\nu}=g_{\mu\nu}-\eta_{\mu\nu}
\ee
and $\eta^{\mu\nu}$ is the Minkowskian metric given by (\ref{minkowski}).
It should be pointed out that all indices on $h_{\mu\nu}$ and/or
$\partial/\partial x_{\mu}$ are raised or lowered with the use of the Minkowskian metric.
Additionally, the  energy-momentum complex of Weinberg is symmetric in its indices
\be
\tau^{\mu\nu}=\tau^{\nu\mu}
\ee
while the superpotential $Q^{\rho\nu\lambda}$ is antisymmetric in its first two indices
\be
Q^{\rho\nu\lambda}=-Q^{\nu\rho\lambda}
\hspace{1ex}.
\ee
The energy and momentum in Weinberg's
prescription for a three-dimensional background are given by
\be
P^{\nu}=\int\int\int
\tau^{\nu t}dx^{1}dx^{2}
\label{weinmomentum}
\ee
and specifically the energy of the physical system in a three-dimensional
background is
\be
E=\int\int\int \tau^{tt}dx^{1}dx^{2}\hspace{1ex}.
\label{weinenergy}
\ee
It should be noted again that the calculations in the Weinberg's prescription
as in all three afore-mentioned prescriptions have to be restricted
to the use of quasi-Cartesian coordinates.
\par\noindent
Since our aim is to evaluate the energy and momentum distributions in the non-static ($2+1$)-dimensional black hole
background described by the line element (\ref{metric1}), we firstly evaluate the Weinberg's
superpotentials. There are sixteen  nonvanishing superpotentials in the Weinberg's prescription
for the specific gravitational background given by (\ref{metric1}) but the required ones are the following
\bea
Q^{ttt}&=&0\,,\nn\\
Q^{xtt}&=&\frac{x\left(1+\Lambda\left(x^{2}+y^{2}\right)+m\right)}{r^2}\,,\nn\\
Q^{ytt}&=&\frac{y\left(1+\Lambda\left(x^{2}+y^{2}\right)+m\right)}{r^2}\,,\nn\\
Q^{txt}&=&-\frac{x\left(1+\Lambda\left(x^{2}+y^{2}\right)+m\right)}{r^2}\,,\nn\\
Q^{xxt}&=&0\,\label{weinsuper}\\
Q^{yxt}&=&0\,,\nn\\
Q^{tyt}&=&-\frac{y\left(1+\Lambda\left(x^{2}+y^{2}\right)+m\right)}{r^2}\,,\nn\\
Q^{xyt}&=&0\,,\nn\\
Q^{yyt}&=&0\hspace{1ex}.\nn
\eea
The mass parameter $m=m(u)$ in terms of the quasi-Cartesian
coordinates is of the form
\be
m(u)=m(t,\sqrt{x^{2}+y^{2}})\hspace{1ex}.
\ee
Substituting the Weinberg's superpotentials (\ref{weinsuper}) into equation (\ref{weintheta}),
the energy density distribution takes the form
\be
\tau^{tt}=\frac{m'}{2\kappa r}+\frac{\Lambda}{\kappa}
\label{weindensity}
\ee
while the momentum density distributions are given by
\bea
\tau^{xt}&=&-\frac{x\dot{m}}{2\kappa r^{2}}\nn\\
\tau^{yt}&=&-\frac{y\dot{m}}{2\kappa r^{2}}\hspace{1ex}.
\label{weinmomendensity}
\eea
Thus, we substitute expressions (\ref{weindensity}) and (\ref{weinmomendensity}) into equations
(\ref{weinenergy}) and (\ref{weinmomentum}), respectively, and the energy and momentum distributions
associated with the non-static ($2+1$)-dimensional black hole under study, which are contained in
a ``sphere''  of radius $r_{0}$,  are given by
\bea
E(r_{0})&=&\frac{\pi}{\kappa}\left(m + \Lambda r^{2}_{0}\right)\nn\\
P_{x}&=&0\label{weindistrib}\\
P_{y}&=&0\hspace{1ex}.\nn
\eea
It is evident again that we have again derived exactly the same energy and momentum distributions
associated with the the non-static ($2+1$)-dimensional black hole as in the cases of the
Einstein's, Landau-Lifshitz's and Papapetrou's prescriptions.
The comment concerning the neutral test particle and the effective gravitational
mass made in Section 1 also holds here.
\par\noindent
Furthermore, the  energy-momentum complex of Weinberg
as formulated in the ($2+1$)-dimensional black hole under study, satisfies the local
conservation laws
\be
\frac{\partial }{\partial x^{\nu}}\,\tau^{\mu\nu}=0
\hspace{1ex}.
\ee
\section{Conclusions}
In this work we have explicitly evaluated the energy and momentum distributions contained in a ``sphere"
of fixed radius of a ($2+1$)-dimensional
black hole. The gravitational background under consideration is an exact solution of Einstein's field
equations in the presence of a cosmological constant $\Lambda$ and a null fluid. Thus, it is regarded as
a ($2+1$)-dimensional black hole with an outflux null radiation. The energy and momentum distributions
are evaluated using four different energy-momentum complexes, specifically these are the energy-momentum
complexes of Einstein, Landau-Lifshitz, Papapetrou and Weinberg.
All four prescriptions give exactly the same energy and
momentum distributions for the specific gravitational background. Consequently, the results obtained here
support the claim that for a given gravitational background, different energy-momentum complexes can
give exactly the same results in three dimensions\footnote{Recently, it was proven by the
author that this is not the case for the two-dimensional stringy black hole backgrounds \cite{elias}.
Particularly, it was shown that Einstein's energy-momentum complex provides meaningful physical results, i.e.
energy and momentum distributions, while M{\o}ller's energy-momentum complex fails to do so.}
as they do in four dimensions.
\par
The ($2+1$)-dimensional black hole solution under study was found by Virbhadra.
He utilized the Landau-Lifshitz energy-momentum
complex in order to evaluate the energy and momentum distributions in the case where the cosmological constant
is zero. The results presented here can be viewed as a generalization of the ones derived by Virbhadra since
our results in the limit of zero cosmological constant reproduce those obtained by Virbhadra.
\par
It should also be pointed out that the energy distribution derived here
can be regarded as the effective gravitational mass
experienced by a neutral test particle placed in the
($2+1$)-dimensional black hole background under consideration.
We have shown that in the case of a negative cosmological constant
it is possible for the neutral test particle although in a gravitational background
to move freely when a condition is fulfilled.
\par
It is also noteworthy to observe that when the mass function of the ($2+1$)-dimensional black hole under
investigation is set constant, the resultant gravitational background is  the spinless BTZ black hole.
Accordingly, one can derive from our results presented here the energy and momentum distributions associated with
the spinless ($2+1$)-dimensional BTZ black hole just by setting the mass function constant. Furthermore, it seems
quite interesting to calculate the energy and momentum distributions associated with the BTZ black hole by
utilizing the M{\o}ller's energy-momentum complex and to compare these results with the corresponding results
derived here. Additionally, it will be also of some interest to investigate the contributions, if any, of the
angular momentum of the BTZ black hole to the energy and momentum distributions. We hope to return to these issues
in a future work.
\par
Finally, since the (2+1)-dimensional BTZ black hole is an asymptotically Anti-de-Sitter spacetime (AAdS), it would
be an oversight not to mention that in the last years due to the AdS/CFT correspondence there has been much
progress in obtaining finite stress energy tensors of AAdS spacetimes\footnote{For a short review see
\cite{sken}}. The gravitational stress energy tensor is in general infinite due to the infinite volume of the
spacetime. In order to find a meaningful definition of gravitational energy one should subtract the divergences.
The proposed prescriptions so far were ad hoc in the sense that one has to embed the boundary in some reference
spacetime. The important drawback of this method is that it is not always possible to find the suitable reference
spacetime. Skenderis and collaborators\footnote{Right after the first work of  Henningson and Skenderis
\cite{hen1}, Nojiri and Odintsov \cite{nojiri1} calculated a finite gravitational stress energy tensor for an AAdS
space where the dual conformal field theory is dilaton coupled. Furthermore, Nojiri and Odintson \cite{nojiri2},
and Ogushi \cite{nojiri3} found well-defined gravitational stress energy tensors for AAdS spacetimes in the
framework of higher derivative gravity and of gauged supergravity with single dilaton respectively. }
\cite{hen1,hen2,skenderis,haro}, and also Balasubramanian and Kraus \cite{bal}, described and implemented a new
method  which provides an intrinsic definition of the gravitational stress energy tensor. The computations are
universal in the sense that apply to all AAdS spacetimes. Therefore, it is nowadays right to state that the issue
of the gravitational stress energy tensor for any AAdS spacetime has been thoroughly understood.
\section*{Acknowledgements}
The author would like to thank  Associate Professor T. Christodoulakis and M.Sc. G.O. Papadopoulos for useful
discussions, and also Professor R.B. Mann for useful correspondence. The author is grateful to Professor K.
Skenderis for his useful suggestions and enlightening comments on the manuscript.
 This work has been supported by the
European Research and Training Network ``EUROGRID-Discrete Random
Geometries: from Solid State Physics to Quantum Gravity"
(HPRN-CT-1999-00161).

\end{document}